\documentclass[a4paper,twoside,11pt]{article}
\usepackage{a4}
\usepackage[english]{babel}
\usepackage{multirow}
\usepackage{graphicx}
\usepackage{exscale}
\usepackage{float}
\usepackage{here}
\usepackage{textcomp}
\usepackage{epsfig}
\begin{document}

\title{Relativistic Heavy Ion Physics \\
{\large (Accepted for publication in Landolt-B\"ornstein I/23A)}}

\date{}

\maketitle

\section{Introduction}
Reinhard Stock\\
stock@ikf.uni-frankfurt.de 

\subsection{Overview}

Quantum Chromodynamics (QCD), the gauge theory of strong interaction, is firmly rooted within the Standard Model of elementary interaction. The elementary constituents of QCD, quarks and gluons that carry the color charge field, have all been observed, in a wealth of by now "classical" experiments in particle physics, such as deep inelastic electron-proton scattering or jet production in electron-positron annihilation and proton-antiproton collisions, at up to TeV center of mass energies. Common to all such observations of partons is an extremely high spatial resolution scale (provided by very high momentum transfer), that allows to recognize a universality of parton interaction at vanishingly small distance: irrespective of their attached fields, and large scale environment. The "running" QCD coupling constant becomes small enough to treat partonic interactions in a perturbative QCD framework amenable to exact solutions (analogous to QED). \\

On the contrary, QCD at modest resolution remains one of the open sectors of the Standard Model. In particular, the confinement-deconfinement transition between hadrons and partons and, more generally, the transition
between hadronic and partonic extended matter have remained un-addressed in the course of particle physics progress. These are of fundamental importance toward the understanding of the primordial cosmological expansion that passed through the QCD color neutralization phase transition to hadrons at about 5 microseconds, giving rise to all ponderable matter in the present universe. One has to address extended continuum QCD matter at extremely high energy density. At hadronization, the energy density of cosmological matter amounts to about 1 GeV per cubic-Fermi ($fm^{3}$), corresponding to about 2$\times$10$^{18}$ kg per cubic meter, and the strong interaction coupling constant is high, almost unity, and thus deep into the non-perturbative sector of QCD. \\

Likewise, neutron star interior matter, or matter dynamics in neutron star mergers (that give rise to heavy nuclei in the interstellar medium and in planets) require knowledge of high density hadronic, or perhaps even quark matter. In addition to the QCD confinement-deconfinement phase transformation (believed to result from gluonic screening of the long range part of color forces), a further characteristic QCD phase transition is involved in hadronic matter close to the critical energy density: the restoration of chiral symmetry in QCD matter. This is an invariance of the QCD Lagrangian, at least for the near-massless light quarks that constitute all cold matter in the universe. It is spontaneously broken in the transition from partons to massive hadrons (this breaking being the origin of allmost the entire hadron mass). Hadron mass is the consequence of non-perturbative vacuum condensates of QCD, which are expected to "melt" as matter approaches the critical conditions. Non perturbative QCD can be numerically approached by solutions on the dicretized space-time lattice, and the finite temperature sector of lattice QCD theory is under intense recent development. \\

What is required in the research field of matter under the governance of the strong interaction is the PHASE DIAGRAM of extended QCD matter, and the EQUATION OF STATE (EOS) governing the relationship of pressure to density, in each of its characteristic domains in density and temperature. Traditional nuclear physics could only offer insight into the ground state of extended QCD matter, and traditional particle physics has dealt essentially only with near-groundstate hadrons and their intrinsic structure. Both fields have merged in RELATIVISTIC HEAVY ION PHYSICS, the topic of this Volume: the study of collisions of heavy nuclear projectiles at relativistic energy. In such collisions an initial dynamics of compression and heating converts the incident, cold nuclear ground state matter into a "fireball" of hadronic or partonic matter, thus populating the QCD matter phase diagram, and notably the deconfined state of a QUARK-GLUON-PLASMA, predicted by lattice QCD to exist over a wide domain of temperature and density. As it turns out the energy available from current synchrotron (CERN SPS) or collider facilities (RHIC at BNL, LHC at CERN) suffices to reach plasma temperatures of up to 1 GeV, i.e. far beyond the QCD phase transition critical temperature, of about 170 MeV.\\

Overall, theoretical studies of QCD in the non-perturbative regimes indicate that QCD matter has a rich phase structure. The phase diagram can be parametrized by the grand canonical variables, temperature T and baryochemical potential $\mu{B}$. Based on the phase diagram, as elucidated by relativistic nuclear collision studies, we obtain perspectives on how the vacuum structure of the early universe evolved in extremely high T states after the Big Bang, as well as what happens in states of extreme baryon density, at the core of neutron stars, and in their merger collisions. Above the deconfinement transition line of the phase diagram we confront a novel partonic continuum state, the quark gluon-plasma(QGP). It turns out to feature strikingly unexpected features, behaving as a strongly coupled liquid state with almost vanishing shear viscosity. In fact this plasma state may turn out to be the first experimentally accessible realization of string theory as it appears that the strongly coupled liquid state lends itself to a calculable framework found in the 5-dimensional AdS/CFT theory. A comprehensive and quantitative understanding of the QCD phase diagram is the most important subject in modern nuclear physics. \\

\subsection{History}

The research field of Relativistic Heavy Ion Collisions was born in the late 1960's, from a coincidence of questions arising in astrophysics (neutron star interior matter, supernova dynamics, early stages in the cosmological evolution) and in fundamental nuclear/hadronic physics (extended nuclear matter and its collective properties, excited hadronic matter and its limits of existence). Generalizing such aspects we see that a description was sought of the phase diagram of strongly interacting matter, in the variables of temperature (big bang and hadronic matter limiting temperature) and matter density (notably invited by the extreme baryon density expected in the neutron star interior). In any corner of this phase diagram the macroscopic statics and dynamics would be determined by an appropriate equation of state (EOS), relating pressure to density and temperature of strongly interacting matter. \\ 

The first employ of the EOS concept was made in the hydrostatic equilibrium model for neutron star density profiles and stability, by Oppenheimer and Volkov~\cite{1}. We are dealing with truely macroscopic, if not gigantic nuclear matter extensions here, but it is noteworthy to recall that, by 1960, nuclear physics had arrived at the realization that even the baryonic matter inside heavy nuclei, however small, features a continuous, quasi macroscopic density distribution, with gradients large as compared to the elementary constituent nucleon force range, and featuring collective, quasi macroscopic modes of excitation. From among those, the observation of the collective giant monopole density vibration mode of heavy nuclei~\cite{2} had yielded first information concerning the pressure to density relation (the EOS) of extended nuclear matter, albeit in an extremely narrow density window only, centered at the nuclear matter ground state density, $\rho_{0}$ = 0.15 baryons per $fm^{3}$. In fact, all other nuclear reaction studies performed in the preceding 50 years of nuclear physics had, likewise, never involved bulk nuclear density changes exceeding the percent level, owing to the fact that the employed accelerators yielded projectile energies of below about 20 MeV, commensurable to the first excitation modes of ground state nuclear matter. In marked contrast, neutron star interior densities were then expected to range beyond five times nuclear ground state density. \\ 

Reaching such densities in the laboratory requires an input of about 100 MeV per nucleon into nuclear matter compressional potential energy, and it was the relativistic shock compression model pioneered by Greiner and collaborators~\cite{3} that first promised just that. The idea was to bombard two heavy nuclei head-on at "relativistic" energy, as defined by the requirement that the relative interpenetration velocity of the two nuclear density profiles be well in excess of the nuclear sound velocity, as estimated from the giant monopole resonance energy, such that a relativistic Mach shock flow phenomenon generated "fireballs" of (excited) hadronic matter, compressed to densities exceeding 2$\gamma \rho_{0}$, with gamma the Lorentz-Factor of the nuclear projectile in the overall center of mass frame. This consideration suggested a projectile energy (in a fixed target experiment) in the 1 to 2 GeV per nucleon range. The basic underlying hypothesis was that nuclear matter in collisions of heavy nuclei was both extended and interactive enough to allow for a hydrodynamic description assumptions that we know now to be fully satisfied. \\ 

Concurrently, acceleration of nuclear projectiles to the required energies was successfully accomplished in Synchrotron laboratories (Berkeley, Dubna and Princeton). As acceleration of heavy nuclei implies a substantial effort of creating and maintaining high levels of projectile ionization, the new field became known as "Relativistic Heavy Ion Physics". Today, relativistic superconducting cyclotrons, and ranges of synchrotrons, provide for any desired nuclear projectile and energy, ranging up to the unfathomable total energy of about 1000 TeV, to be reached in Lead-on-Lead collisions at the CERN LHC facility in 2010. \\ 

The goal, of delineating the hadronic matter equation of state, has indeed been matched in nuclear collision studies in the GeV-domain (see "The Quest for the Nuclear Equation of State" in this Volume), albeit after more than two decades of systematic effort. The obstacle in the path toward the zero temperature EOS, as relevant to neutron star structure, and neutron star mergers: the initial T = 0 matter of nuclear projectiles gets compressed but also heated, to beyond 100 MeV, in such collisions. The collisional reaction dynamics is thus sensitive to isothermes of thermally excited, and compressed matter, and the T = 0 EOS could only be derived in a semi-empirical manner, involving relativistic theory hadron transport occuring at the microscopic level, which required a substantial corresponding theoretical effort and innovation. Quite in general strong interaction theory has received, and reacted to, quite a substantial stimulus, over decades, as emanating from a surprisingly multi-faceted development of highly "provocative" experimental observations. We shall turn to prominent examples below. \\ 

The QCD-revolution of strong interaction physics, occuring in the 1970's, (the development of the non-abelian gauge field theory, Quantum Chromodynamics, a sector of what concurrently became known as the "Standard Model" of elementary interaction) then provided the field of relativistic nuclear collision studies with an unprecedented uplift of scope, to which it is responding until today. With the realization of partons, i.e. quarks and gluons, as the elementary carriers of the "colour" charge of the strong interaction force field, the study extended strongly interacting matter faced a new goal: to look $beyond$ the limits of hadronic matter stability. Such limits had resulted from Hagedorn~\cite{4} exploring the "statistical bootstrap model" of hadronic/resonance/fireball matter, which had resulted in a limiting energy density of about one GeV per $fm^{3}$, with corresponding temperature in the vicinity of T = 160 - 170 MeV, the famous "Hagedorn limiting temperature". QCD now implied a partonic matter phase beyond these limits, the "Quark-Gluon Plasma"(QGP), as it was baptized by Shuryak~\cite{5}. This implies the consideration of a phase transition occuring, with energy density transcending the parton-hadron phase boundary (predicted at about 1GeV/$fm^{3}$ by Hagedorn), such that partons confined in colour neutral hadrons become the effective degrees of freedom of  colour-conducting QCD-matter. This appeared to be also intuitively plausible because, above the critical density, a single hadronic volume would contain more than one thermal gluon, causing force screening and resulting in deconfinement. \\ 

Immediately, the cosmological temperature/density evolution came into view. Based on considerations of QCD deconfinement by Collins, Perry, Cabibbo and Parisi~\cite{6}, supposed to result from the falloff of the QCD coupling constant at extremely high temperatures, in the multi-GeV domain (a phenomenon called asymptotic freedom), Weinberg famously argued in 1976 that "quarks were close enough together in the early universe so that they did not feel these (binding) forces, and could behave like free particles"~\cite{7}. Seen in retrospect, the initial qualitative ideas concerning deconfinement to a Quark-Gluon Plasma did range from dissociation by colour "Debye" screening, to thermal dissociation, and dissolution of partonic bound states by asymptotic freedom. All these pictures are leaning on a perturbative treatment of QCD. However, also expecting the partonic phase to set in just above the Hagedorn limiting hadronic temperature, at T about 160 to 170 MeV, and at the corresponding  resolution scale of dimension 1 Fermi, deconfinement clearly was a non perturbative process, and the confinement/deconfinement transition remained a major open problem of QCD, as did the order of the phase transformation. \\ 

It became clear that relativistic nuclear collisions at higher than GeV energies offered the possibility to dive deeply into the QCD matter diagram, transcending the critical QCD energy density. In 1974 T.D. Lee was the first to formulate an appropriate, non perturbative vision of QGP creation in nuclear collisions where "the non-perturbative vacuum condensates could be melted down...by distributing high energy or high nucleon density over a relatively large volume"~\cite{8}. Note that hadrons represent such non-perturbative condensates (chiral symmetry breaking excitations of the vacuum). This vision, of entering the phase transition in a process of chiral symmetry restoration occuring near a critical temperature $T_{c}$, and probably coinciding with the Hagedorn temperature, was substantiated, a decade later, by first calculations employing a lattice discretization scheme for non-perturbative analysis of QCD matter at finite temperature~\cite{9}, which indeed exhibited a sharp melting transition of the chiral condensate (the vacuum expectation value of the $<\bar{\psi},\psi>$
term in the QCD Lagrangian) at about 170 MeV~\cite{10}. \\ 

The emerging goal of relativistic nuclear collision study was, thus, to locate this transition, elaborate its properties, and gain insight into the detailed nature of the deconfined QCD phase. Required beam energies turned out to be upward of about 10 GeV per nucleon pair in the CM frame, i.e. $\sqrt{s}$ $>$10 GeV, and various experimental programs have been carried out, and are being prepared, at the CERN SPS (up to about 20 GeV), at the BNL RHIC collider (up to 200 GeV), and finally reaching 5.5 TeV at the CERN LHC in 2009. This Volume attempts an overview of the most outstanding results and emerging perspectives. \\ 

\subsection{The QCD Phase Diagram}

\begin{figure}[h]   
\begin{center}
\includegraphics[scale=0.5]{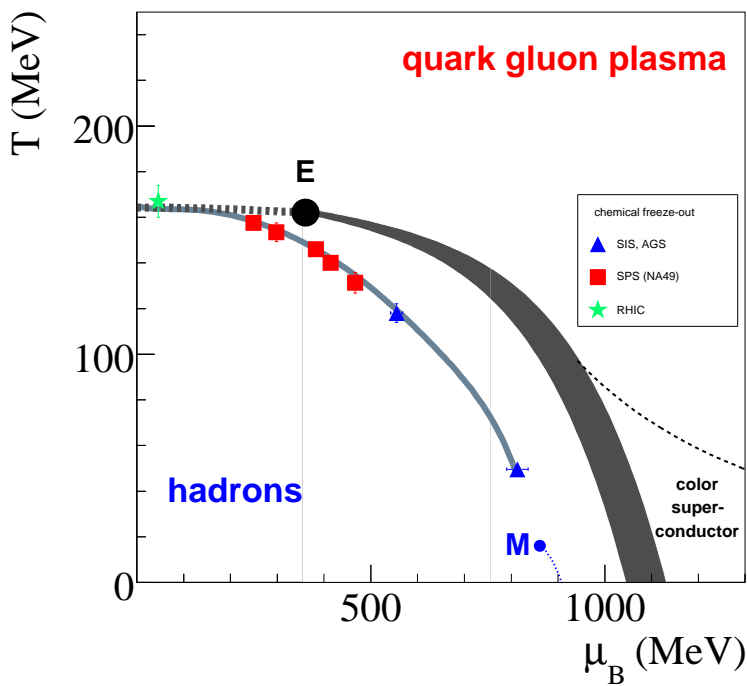}
\caption{Sketch of the QCD matter phase diagram in the plane of temperature $T$ and baryo-chemical potential $\mu_B$. The parton-hadron phase transition line from lattice QCD  \cite{11,12,13} ends in a critical point $E$. A cross-over transition occurs at smaller $\mu_B$. Also shown are the points of hadro-chemical freeze-out from the grand canonical statistical model.}
\label{fig:Figure1}
\end{center}
\end{figure}

QCD confinement-deconfinement transitions are by no means limited to the domain of the phase diagram that is relevant to cosmological expansion dynamics prior to about 5 microseconds (the time of the hadronization transition), where a vanishingly small excess of baryon over antibaryon density implies near zero baryo-chemical potential $\mu_{B}$. In fact, modern QCD suggests~\cite{11,12,13} a detailed phase diagram, with various forms of strongly interacting matter and states, that we sketch in Fig.~\ref{fig:Figure1}. It is presented in the plane of temperature T and baryochemical potential $\mu_{B}$. We are thus employing the terminology of the grand canonical Gibbs ensemble that describes an extended volume $V$ of partonic or hadronic matter at temperature $T$. In it, total particle number is not conserved at relativistic energy, due to particle production-annihilation processes occurring at the microscopic level. However, the probability distributions (partition functions) describing the relative particle species abundances have to respect the presence of certain, to be conserved net quantum numbers ($i$), notably non-zero net baryon number and zero net strangeness and charm. Their global conservation is achieved by a thermodynamic trick, adding to the system Lagrangian a so-called Lagrange multiplier term, for each of such quantum number conservation tasks. This procedure enters a "chemical potential" $\mu_i$ that modifies the partition function via an extra term $\exp{\left(-\mu_i/T\right)}$ occuring in the phase space integral. It modifies the canonical "punishment factor" ($\exp{\left(-E/T\right)}$), where E is the total particle energy in vacuum, to arrive at an analogous grand canonical factor for the extended medium, of $\exp{\left(-E/T - \mu_i/T\right)}$. This concept is of prime importance for a description of the state of matter created in heavy nuclear collisions, where net-baryon number (valence quarks) carrying objects are considered. The same applies to the matter in the interior of neutron stars. Note that $\mu_B$ is high at low energies of collisions creating a matter fireball. In a head-on collision of two mass 200 nuclei at $\sqrt{s}$ =15 GeV the fireball contains about equal numbers of newly created quark-antiquark pairs (of zero net baryon number), and of initial valence quarks. The accomodation of the latter, into created hadronic species, thus requires a formidable redistribution task of net baryon number, reflecting in a high value of $\mu_B$. Conversely, at LHC energy (5.5TeV for Pb+Pb collisions), the initial valence quarks constitute a mere 5\% fraction of the total quark density, correspondingly requiring a small value of $\mu_B$. In the extreme, big bang matter evolves toward hadronization (at $T$=170 MeV) featuring a quark over antiquark density excess of $10^{-9}$ only, resulting in $\mu_B \approx 0$. The limits of existence of the hadronic phase are not only reached by temperature increase, to the so-called Hagedorn value $T_H$ (which coincides with $T_{crit}$ at $\mu_B \rightarrow 0$), but also by density increase to $\varrho > (5-10)\: \varrho_0$: ''cold compression'' beyond the nuclear matter ground state baryon density $\varrho_0$ of about 0.16 $B/fm^3$. We are talking about the deep interior sections of neutron stars or about neutron star mergers \cite{14,15,16}, at low $T$ but high $\mu_{B}$. \\ 

A sketch of the present view of the QCD phase diagram \cite{11,12,13} is given in Fig.~\ref{fig:Figure1}. It is dominated by the parton-hadron phase transition line that interpolates smoothly between the extremes of predominant matter heating (high $T$, low $\mu_B$) and predominant matter compression ($T \rightarrow 0, \: \mu_B > 1 \: GeV$). Onward in $T$ from the latter conditions, the transition is expected to be of first order \cite{17} until a critical point of QCD matter is reached at $200 \le \mu_B \: (E)\: \le 500 \: MeV$. The relatively large position uncertainty reflects the preliminary character of Lattice QCD calculations at finite $\mu_B$ \cite{11,12,13}. Onward from the critical point, E, the phase transformation at lower $\mu_B$ is a cross-over\cite{13,18}, thus also including the case of primordial cosmological expansion. 
This would finally rule out former ideas, based on the picture of a violent first order''explosive'' cosmological hadronization phase transition, that might have caused non-homogeneous conditions, prevailing during early nucleo-synthesis \cite{19}, and fluctuations of global matter distribution density that could have served as seedlings of galactic cluster formation \cite{20}. However, it needs to be stressed that the conjectured order of phase transformation, occuring along the parton - hadron phase boundary line, has not been unambiguously
confirmed by experiment, as of now. \\ 

On the other hand, the {\it position} of the QCD phase boundary at low $\mu_B$ has, in fact, been located by the hadronization points in the $T, \: \mu_B$ plane that are also illustrated in Fig.~\ref{fig:Figure1}. They are obtained from statistical model analysis \cite{21} of the various hadron multiplicities created in nucleus-nucleus collisions, which results in a [$T, \: \mu_B$] determination at each incident energy, which ranges from SIS via AGS and SPS to RHIC energies, i.e. $3\le \sqrt{s} \le 200 \: GeV$. Toward low $\mu_B$ these hadronic freeze-out points merge with the lattice QCD parton-hadron coexistence line: hadron formation coincides with hadronic species freeze-out, at high $\sqrt{s}$. These points also indicate the $\mu_B$ domain of the phase diagram which is accessible to relativistic nuclear collisions. The domain at $\mu_B \ge 1.5 \: GeV$ which is predicted to be in a further new phase of QCD featuring color-flavor locking and color superconductivity \cite{22} will probably be accessible only to astrophysical observation. \\ 

In Fig.~\ref{fig:Figure1} we are representing states of QCD matter in thermodynamic equilibrium. What is the relation of such states, e.g. a Quark-Gluon Plasma at some T and $\mu_{B}$, to the dynamics of relativistic nuclear collisions? A detailed answer can be only given based on a microscopic transport description of the dynamical evolution, and many of the articles in this Volume address the occurence, and generation, of local or even global thermal and/or chemical equilibrium conditions. As an example, the points of hadronic freeze-out in Fig.~\ref{fig:Figure1} refer to the observation of perfect hadro-chemical equilibrium among the created species, the derived (T,$\mu_{B}$) value~\cite{21} thus legitimately appearing in the phase diagram. We may add that, in general, the collisional reaction volume of head-on collisions of heavy nuclei is of dimension 10 fm in space and time whereas the typical microscopic extension, and relaxation time scale of non-perturbative QCD objects is of order 1 fm (the confinemant scale). The A+A collision fireball size thus exceeds, by far, the elementary dimensions of microscopic strong interaction dynamics. Further, one can only get with the help of detailed microscopic models. \\ 

\subsection{Physics Observables}

One can order the various physics observables, that have been developed in this field and are described in this Volume, in sequence of their origin, from successive stages that characterize the overall dynamical evolution of relativistic collisions at high $\sqrt{s}$. In rough outline this evolution can be seen to proceed in three major steps. An initial period of matter compression and heating occurs in the course of interpenetration of the projectile and target baryon density distributions. Inelastic processes occuring at the microscopic level convert initial beam longitudinal energy to new internal and transverse degrees of freedom, by breaking up the initial baryon structure functions. Their partons thus acquire virtual mass, populating transverse phase space in the course of inelastic perturbative QCD shower multiplication. This stage should be far from thermal equilibrium, initially. However, in step two, inelastic interaction between the two arising parton fields (opposing each other in longitudinal phase space) should lead to a pile-up of partonic energy density centered at mid-rapidity (the longitudinal coordinate of the overall center of mass). Due to this mutual stopping down of the initial target and projectile parton fragmentation showers, and from the concurrent decrease of parton virtuality (with decreasing average square
momentum transfer $Q^2$) there results a slowdown of the time scales governing the dynamical evolution. Equilibrium could be approached here, the system ''lands'' on the $T, \: \mu$ plane of Fig.~\ref{fig:Figure1}, at temperatures of about 300 and $200 \: MeV$ at top RHIC and top SPS energy, respectively. The third step, system expansion and decay, thus occurs from well above the QCD parton-hadron boundary line. Hadrons and hadronic resonances then form, which decouple swiftly from further inelastic transmutation so that their yield ratios become stationary (''frozen-out''). These freeze-out points are included in Fig~\ref{fig:Figure1} for various $\sqrt{s}$. A final expansion period dilutes the system to a degree such that strong interaction ceases all
together. \\ 

In order to verify in detail this qualitative overall model, and to ascertain the existence (and to study the properties) of the different states of QCD that are populated in sequence, one seeks observable physics quantities that convey information imprinted during distinct stages of the dynamical evolution, and ''freezing-out'' without significant obliteration by subsequent stages. Clearly, the dynamical formation of a local or even more extended near-equilibrium quark-gluon plasma medium is of foremost interest, but this process belongs to the very early evolution, right after the termination of the primordial perturbative shower evolution period, at about 0.3 to 0.5 fm/c. The parton density and the microscopic rescattering collision frequency are maximal during this period, but the likelyhood to receive unobliterated signals of such early evolution is evidently minute. Nevertheless this problem has been overcome in a remarkable quest for suitable early-time observables, notably the dissolution of primordially formed charm-anticharm quark pairs that would evolve into charmonium states like $J/\Psi$, in elementary collisions, but get obliterated in the course of their traversal of hot and dense early fireball matter~\cite{23}, resulting in a suppression of the eventually observed charmonium production rate. Likewise, the high $p_{t}$ partons emerging in the course of primordial, perturbative first parton shower and jet production experience a dramatic dampening characteristic of early medium opacity conditions, leading to a general, well observable high $p_{t}$ hadron suppression. It thus turned out that the primordial, perturbative mechanisms provide for several "tracer probes" co-travelling with the ongoing early dynamical evolution as diagnostic agents. These observables yield information, almost similar to what would be provided by a fictitious deep inelastic scattering experiment with partonic fireballs as a target. Moreover, it has turned out that early collective partonic flow modes of relativistic hydrodynamic matter get formed in the collisional volume, surviving later expansion stages, including hadronization, and providing for information on e.g. medium viscosity~\cite{24} and equation of state~\cite{25}, at times below and at 1 fm/c. \\

This focus on primordial time evolution came about in the course of experimental progress, from SPS to RHIC energies. It is, basically, a straight forward consequence of increasing time resolution, brought about by the collider technique that extended the energy range toward $\sqrt{s}$ = 200GeV. Consider the duration of primordial interpenetration of the projectile-target nuclear density distributions, t = $2R/\gamma$, where R is the nuclear radius. At top SPS energy t = 1.6 fm/c whereas, at RHIC top energy, t = 0.14 fm/c. Evidently, the critical early time interval below about 1 fm/c can not be properly resolved at top SPS energy: at time as "late" as about 1.5 fm/c, initial spatial layers of the interaction volume are well past all primordial interaction stages, whereas nucleons at the far end layers of the density distributions still enter their first interaction. Thus, there exists no global, time-synchronized interacting system until later times, of about 2 fm/c and beyond, i.e. just before the hadronization transition sets in: SPS physics captures a partonic system at densities of about 2-3 GeV per $fm^{3}$, in the vicinity of the parton-hadron coexistence line predicted by lattice QCD. Whereas, at top RHIC energy, t = 0.14 fm/c, and synchronization of a global high density QCD matter evolutional trajectory (including a local approach toward partonic equilibrium) may be accomplished at times as low as 0.5 fm/c, thus enabeling the definition of the above-mentioned early-time observables. It is, thus, not surprising that the "elliptic flow" signal, resulting from collective primordial partonic density/pressure gradients, stays relatively small up to top SPS energy but reaches the "hydrodynamic limit" ( i.e. the elliptic flow magnitude predicted by parton hydrodynamics) at top RHIC energy. \\ 

At future LHC energy the interpenetration time is practically zero. Thus even the primordial, perturbative QCD parton shower multiplication phase, occuring at times below about 0.2 fm/c, will now occur in a globally synchronized, extended collisional volume. This gives rise to the expectation~\cite{26} that effects of QCD colour saturation become essential in the primordial evolution: at the relatively low momentum transfers, corresponding to "soft" bulk hadron production, individual colour charges in the projectile-target transverse parton density profiles can not be spatially resoved in the dynamics. A new version of QCD is beeing called for: established QCD "DGLAP" evolution treats only interactions of elementary unit colour charge~\cite{27}, unlike classical Maxwell theory that deals with arbitrary charge Z force fields (recall the superposition principle). What one looks for, here, is something like a "classical limit" of QCD interaction, which has been called a "colour glass condensate" theory~\cite{28}. \\ 

On the path from top SPS to top RHIC energy, exploiting the physics observables that refer to early time stages in the dynamical evolution has resulted in substantial first contributions to our view of the deconfined Quark-Gluon Plasma state,  expected to exist above the parton-hadron coexistence line in Fig.~\ref{fig:Figure1}. A multiply cross-connected web of theoretical inferrences from the RHIC early time signals (in part also oftentimes called "hard probes" because their initial "tracer" partons emerge from a hard , high $Q^{2}$ process, described in perturbative QCD) leads to the conclusion that the initial evolutionary stages of RHIC Au+Au collisions reach QCD matter at about 10 times higher than the critical density ( $\epsilon_{c}$ = 1GeV/$fm^{3}$), and temperatures of 300-400 MeV. This matter, most remarkably, appears to be $very$ $much$ $unlike$ a free gas of weakly coupled partons, rather behaving like a near-ideal liquid, with a minimal shear viscosity~\cite{29}. As such, however, it is clearly a non-perturbative QCD matter, and a whealth of new focus on the development of npQCD theory has resulted. In particular it was observed~\cite{30} that a "dual" theory might exist for the strongly coupled, non perturbative QCD plasma, which is weakly coupled enabling a quantitative description of viscosity and other transport properties. This dual theory turns out to be a 5-dimensional string theory in anti-de Sitter (AdS) topology. This theory now confronts the alternative QCD lattice theory - both in need, and in evidence, of progress, which results in an unprecedented focus of fundamental interest in this field. \\ 

Concerning the further evolution, the medium conditions governing matter in the direct vicinity of the parton-hadron confinement transition have been illucidated by data gathered at the lower SPS energy , where the dynamical trajectory of A+A collisions can be expected to settle at a "turning point", occuring inbetween the overall compression-expansion cycles, and situated close to the line of QCD phase transformation. Two fundamental QCD symmetry breaking transitions, with falling energy density, are encountered here. Confinement leads to hadron formation as, at the "critical energy density", coloured partons acquire lower free energy via pre-hadronic colour singlet formation~\cite{31}. Concurrently, as lattice QCD results suggest, albeit at zero baryo-chemical potential only~\cite{13}, chiral symmetry breaking leads to a "dressing" of the partons with non-perturbative vacuum excitation "condensate" mass, eventually leading to the observed hadronic mass. The end product of these two transitions emerges as a multitude of hadron and resonance states, produced with a characteristic multiplicity pattern ranging over several orders of magnitude, while the chiral restoration process to hadrons concurs with electromagnetic interaction decay of hadronic, or pre-hadronic resonance states, to observable dilepton final channels. This decay is active throughout the hadronization process. Two principal physics observables result: the hadron-resonance species yield distribution emerging from hadronization, and the dilepton invariant mass spectra that integrate the yield over the entire hadronization period. The former exhibit a striking resemblance to Gibbs grand-canonical equilibrium distributions~\cite{32}, leading to determination of the hadronization temperature and corresponding hadro-chemical potential (recall the entries in Fig.~\ref{fig:Figure1}, of (T,$\mu_{B}$) points for various $\sqrt{s}$). The latter indeed give indications of a gradual broadening and "melting" of the hadronic spectral functions , in the vicinity of the critical temperature~\cite{33}. \\ 

The order of this phase transformation, at non-zero baryochemical potential, has not been unambiguously determined, as of yet. State of the art predictions of lattice QCD show initial success in overcoming the corresponding mathematical obstacles~\cite{11,12,13}, as we have implied by the parton-hadron phase boundary sketched in Fig.~\ref{fig:Figure1}. This line features, in particular, a critical point, at a still rather uncertain position but expected to occur at a rather high baryochemical potential, corresponding to the low end of SPS energies, covered only rather superficially in the CERN SPS Pb-beam program. The critical point would lead to typical critical fluctuations~\cite{34}, and imply an adjacent first order phase transition domain~\cite{35}. Both of these conjectured QCD matter properties should imprint distinct traces onto the dynamical trajectory of A+A collisions- it is just unknown today whether dramatic, or subtle. For appropriate signals one wants to consult observables imprinted during the hadronization stage, e.g. hadronic species equilibrium decoupling freeze-out points, or possible reflections on collective hydrodynamic flow from a "softest point" of the equation of state, reflecting in radial and elliptic flow variables, as well as  pion pair Bose-Einstein interferometry study of collective matter flow toward hadronic decoupling. While first potentially significant data exist~\cite{36} it is clear that this region of the phase diagram deserves intense further study. This is, in fact, the goal of a planned low energy running program at RHIC, and one of the major purposes of the planned FAIR-facility of GSI. \\ 

In summary, from among the Standard Model fundamental interactions and force fields, extended matter architecture arises, most prominently, from the electromagnetic, and the strong interactions (gravitational architecture overwhelmingly evident but still not available from a quantized theory). After decades of nuclear, and hadron physics, the 1970's QCD revolution has, at first, swept aside the non-perturbative sector relevant to QCD as a theory of extended matter, embracing the evidence for a renormalizable gauge field theory resulting from high $Q^{2}$ physics (deep inelastic scattering, jet phenomenology), that triumphantly exploited perturbative QCD, in microscopic processes. However, the present Universe consists, in its "luminous", or "ponderable" extent, of extended non-perturbative QCD matter, ranging from baryonic and nuclear gas stellar concentrations, to extended superfluid dense hadronic, or even quark matter neutron star interiors. Likewise, the primordial cosmological evolution, and its partial resurrection in violent black hole formation, and neutron star merger collisions, features dynamics associated with an appropriate partonic/hadronic equation of state, and occupies certain, characteristic entries into the overall, non perturbative QCD matter phase diagram. The relatively young research field, of relativistic nuclear collision study, has accomplished first outlines of the QCD matter phase diagram, and of extended QCD matter collective interaction dynamics: the subject of this Volume.

\end{document}